# A six-factor asset pricing model

Rahul Roy*, Santhakumar Shijin

*Department of Commerce, School of Management, Pondicherry University, Pondicherry, 605014, India*

## Abstract

The present study introduce the human capital component to the Fama and French five-factor model proposing an equilibrium six-factor asset pricing model. The study employs an aggregate of four sets of portfolios mimicking size and industry with varying dimensions. The first set consists of three set of six portfolios each sorted on size to B/M, size to investment, and size to momentum. The second set comprises of five index portfolios, third, a four-set of twenty-five portfolios each sorted on size to B/M, size to investment, size to profitability, and size to momentum, and the final set constitute thirty industry portfolios. To estimate the parameters of six-factor asset pricing model for the four sets of variant portfolios, we use OLS and Generalized method of moments based robust instrumental variables technique (IVGMM). The results obtained from the relevance, endogeneity, overidentifying restrictions, and the Hausman's specification, tests indicate that the parameter estimates of the six-factor model using IVGMM are robust and performs better than the OLS approach. The human capital component shares equally the predictive power alongside the factors in the framework in explaining the variations in return on portfolios. Furthermore, we assess the *t*-ratio of the human capital component of each IVGMM estimates of the six-factor asset pricing model for the four sets of variant portfolios. The *t*-ratio of the human capital of the eighty-three IVGMM estimates are more than 3.00 with reference to the standard proposed by Harvey et al. (2016). This indicates the empirical success of the six-factor asset-pricing model in explaining the variation in asset returns.

*JEL classification:* G12; C36; J24

*Keywords:* FF portfolio; Human capital; IVGMM approach; Return predictability; Six-factor asset pricing model

## 1. Introduction

The contribution of Black, Jensen, and Scholes (1972), Lintner (1965), Mossin (1966), and Sharpe (1964) in the capital asset pricing literature that develops the understanding between risk-return relationship is impeccable. Since then the empirical tests of asset pricing literature proclaims the inability of capital asset pricing model (CAPM) to price the variations in asset returns. Henceforth, over the last few decades, the asset pricing literature has transformed CAPM into the multi-factor model to price the variation in asset returns.

The prominent model among them is the Fama and French (1993), three-factor model that extends CAPM with two factors relating to book-to-market and size. Successively, the three-factor model of Fama and French (FF) become the benchmark model to price the variation in cross-sectional asset returns.

Lately, Fama and French (2015) proposed a five-factor model by augmenting investment and profitability in their three-factor model to capture the variation in asset returns. FF five-factor model performs remarkably well in explaining the variation in asset returns and simultaneously outperforms the FF three-factor model. Further Chiah, Chai, Zhong, and Li (2016) evidenced that FF five-factor model performs better than the range of competing multi-factor models in explaining the variation in asset return across the global equity markets. Contrastingly, Kubota and Takehara (2018) found that FF five-

* Corresponding author.
  *E-mail addresses:* rahulroy819@gmail.com, rroypu.res@pondiuni.edu.in (R. Roy), shijin.s@gmail.com, shijin.com@pondiuni.edu.in (S. Shijin).

Peer review under responsibility of Borsa İstanbul Anonim Şirketi.

factor model underperforms in explaining the variation in asset returns. Campbell (1996) favored relating value and size strategies with the human capital component. Interestingly, Kim, Kim, and Min (2011) found that human capital component subsumes the predictive power of value and size strategies respectively. Mayers (1972) identified the role of the human capital component in asset return predictability where approximately 75 percent of consumption occurs on labor income encompassing human capital as an invaluable component of aggregate wealth.

Belo, Li, Lin, and Zhao (2017) found strong linkage between asset pricing and labor economics literature where financial variables provide a rich source of information about the significance of labor market dynamics. The empirical results suggest that the characteristic features of labor component largely ignored in capital asset pricing, determine the premiums in the cross-sectional returns. Similarly, Kuehn, Simutin, and Wang (2017) examined the dynamics of interaction between labor and financial markets to figure out the determinants of the cross-section of equity returns. Building on this insight Shijin, Gopalaswamy, and Acharya (2012) found human capital component captures the significant portion of variation in asset returns.

Similarly, the study of Roy and Shijin (2018) examined the dynamics of the human capital component, common factors, and frequently used financial variables in the asset pricing literature. The study found empirical evidence in the international data that market factor and human capital component captures the variation in asset return predictability for the utmost asset. Further, proposes that human capital component subsumes the explanatory power of size and value strategies in return predictability. The fact that human capital component representing the explanatory power of size and value strategies respectively makes it interesting to introduce the component into FF five-factor model. Fama and French (2015) professed that five-factor model was a failure due to its inability to capture the variation in returns on small stocks. Moreover, Racicot & Rentz (2017) examined five-factor model using IVGMM in a panel data framework on FF dataset found that market portfolio is the only significant factor.

Reflecting on the fact that the aggregate wealth consists of financial and human wealth, there exists a great deal of evidence in the asset pricing literature that the framework comprising human capital component yields superior results than the multi-factor models. Hence, building on the theoretical justification that human capital component owes 90 percent of the aggregate wealth (Lustig, Van Nieuwerburgh, & Verdelhan, 2013), we introduce the human capital component to the five-factor model of Fama and French (2015). We aim to propose an equilibrium six-factor asset pricing model by adding the human capital component with the FF five-factor model that consists of the market factor, size, value, profitability, and investment, respectively.

The present study is the first of its kind to propose a six-factor asset pricing model to price the variation in return predictability employing US data. We employ an aggregate of four sets of portfolios that include, firstly, a three set of six portfolios each sorted on the intersections of size- B/M, size-investment, and size-momentum, following Fama and French (1993, 2015). Secondly, a set of five index portfolios, thirdly, a four-set of twenty-five portfolios each sorted on the intersections of, size-B/M, size-investment, size-profitability, and size-momentum, and finally, a set of thirty industry portfolios. To estimate the parameters of the new six-factor asset pricing model for the four sets of variant portfolios, we use Ordinary Least Squares (OLS) and the IVGMM approach. Simultaneously, we perform a weak instrumental variable test (Racicot & Rentz, 2015) to check for the validity of the instruments, tests of overidentifying restrictions (Hansen, 1982; Olea & Pflueger, 2013), and the Hausman test (Hausman, 1978) to check the specification and measurement errors. These robustness tests further validate the economic and statistical significance of the empirical results.

Our approaches lead to the following conclusions; firstly, for the first set of eighteen FF portfolios, the human capital coefficient is significant for thirteen portfolios using OLS while eight are significant using IVGMM approach. For the second set of index portfolios, the human capital is significant for all the five index portfolios using OLS and four using IVGMM approach. Secondly, for the third set of one-hundred FF portfolios, the human capital component is significant for forty-three using OLS and eighty-seven using IVGMM approach. Subsequently, for the fourth set of thirty industry portfolios, the human capital component is significant for three industry portfolios using OLS, while seventeen using IVGMM approach. The relevance, exogeneity, overidentifying restrictions test and the Hausman's specification test results indicate that the parameter estimates of the six-factor asset pricing model using IVGMM approach is robust and performs superior to the OLS. The parameters, human capital, market, and the size factors contain measurement errors while using OLS for the first and second sets of variant portfolios. Further, this suggests that OLS may be overstating the significance of the regressions. Harvey, Liu, and Zhu (2016) argue that unless the $t$-ratio for a factor is more than 3.00, any claimed empirical finding for a factor is likely to be false. For the present study, the $t$-ratio of the human capital of eighty-three IVGMM estimates of the six-factor asset pricing model is more than 3.00. This indicates the empirical success of the new six-factor asset pricing model proposed to explain the variations in asset returns.

Conclusively, these findings contribute to the contemporaneous literature in several ways. Firstly, the present study contributes to the existing literature on multi-factor asset pricing model by proposing a new six-factor asset pricing model with a view to develop a better understanding of risk and return trade-off. Further, the present study successfully unlocks the cause of the failure of FF five-factor asset pricing model (Fama & French, 2015a). Secondly, the present study strengthens considerably the existing literature (Campbell, 1996; Heaton & Lucas, 2000; Lustig et al., 2013) that advocates human capital component owes the greater chunk of an aggregate wealth and is equally an important component apart from financial wealth in returns predictability. Further, the empirical evidence shows that the dynamic human capital component shares the predictive power on par with the rest of the factors in explaining the asset returns. This indicates that the human capital component is an important candidate to be considered while modeling the asset returns in the multi-factor asset pricing framework. Lastly, the empirical evidence that the parameter estimates of the six-factor asset pricing model using IVGMM approach outperforms the conventional OLS owing to specification and measurement errors.

The remainder of this paper is structured as follows. Section 2 describes the data and variables used in the study. This is followed by a brief discussion of the methodology in Section 3. Section 4 explores the empirical results, and both the Section 5 and 6 provides the information about the performance of a six-factor asset pricing model. Section 7 describes the importance of the dynamic human capital in returns predictability, Section 8 presents the summary and discussion, and Section 9 reports the concluding remarks.

## 2. Data and variable definitions

We employ labor income growth rate (LBR) to proxy the human capital component, which is the quarterly salaries and wages component from the earnings report of the firms listed on NYSE, AMEX, and NASDAQ. The salaries and wages data are retrieved from Thomson Reuters Datastream. The aggregate of quarterly salaries and wages is transformed into a monthly frequency to match the frequency with the rest of the variables used in the study assuming labor income growth rate follows a similar pattern in each quarter. RM-RF, SMB, HML, CMA, and RMW mimicking market factor, size, value, investment, and profitability respectively and the definitions are reported in panel A of Table S1 (See the Supplementary Material, available online). We employ three sets of six portfolios each, sorted on the intersections of size-B/M, size-investment, and size-momentum, following Fama and French (1993, 2016). The resulting eighteen portfolios' description has been reported in panel B of Table S1 (See the Supplementary Material, available online). Panel C and D of Table S1 (See the Supplementary Material, available online) shows the description of thirty industry portfolios and five index portfolios, respectively. Further, we employ four set of twenty-five portfolio each, sorted on the intersections of size-B/M, size-investment, size-profitability, and size-momentum, resulting in an aggregate of a one-hundred portfolio of different sorts. The descriptions of portfolio formation are discussed and illustrated in Tables S4, S5, S6, and S7, respectively. Subsequently, the detailed information about FF portfolio formation is available on the Kenneth R. French − Data Library. The required data and the variables employed in the study are retrieved from French (2017) Data library on monthly frequency basis, which covers the period from January 1986 to May 2017 and expressed in terms of USD (US dollar) currency unit.

## 3. Econometric methodology

Fama and French (2015) claims that FF five-factor model is an incomplete equilibrium model due to its inability to capture the variability in return predictability. Moreover, Campbell (1996) professed to relate human capital component with common factors to enhance the statistical and economic performance of the multi-factor asset pricing model. Motivated by this thought process, we proposed novel six-factor model by adding a human capital component with FF five-factor model, to measure the variation in return predictability. Thus, to estimate the parameters of proposed six-factor asset pricing model to price the return on variant portfolios, the time series regression takes the form of OLS,

$$R_{it}^P = \alpha_i + l_i LBR_t + b_i(RM-RF)_t + s_i SMB_t + h_i HML_t + r_i RMW_t + c_i CMA_t + e_{it} \quad (1)$$

In equation (1) $R_{it}^P$ is the return on portfolio $i$ (FF, index, and industry portfolios) for the month $t$. $LBR_t$ is the labor income growth rate for the month $t$. $(RM\text{-}RF)_t$ is the value-weighted market portfolio return minus risk-free rate (91 days Treasury bill rate) for the month $t$. $SMB_t$ is the size strategy for the month $t$, $HML_t$ is the value strategy for the month $t$, $RMW_t$ is the profitability strategy for the month $t$, $CMA_t$ is the investment strategy for the month $t$, and $e_{it}$ is the white noise process.

### 3.1. Instrumental variable generalized method of moments (IVGMM)

The standard IV estimator is the special case of the GMM estimator. The assumptions that the instruments $Z$ are exogenous can be denoted as $E(Z_i u_i) = 0$. The $L$ instruments gives a set of $L$ moments,

$$g_i(\widehat{\beta}) = Z_i' \widehat{u}_i = Z_i'(y_i - X_i\widehat{\beta}) \quad (2.1)$$

where $g_i$ is $L \times 1$. The exogeneity of the instruments means that there are $L$ moment conditions, which will be satisfied at the true value of

$$E(g_i(\beta)) = 0; \quad (2.2)$$

Each of the $L$ moment equations corresponds to a sample moment, and $L$ sample moments takes the form

$$\overline{g}(\widehat{\beta}) = \frac{1}{n}\sum_{i=1}^{n} g_i(\widehat{\beta}) = \frac{1}{n}\sum_{i=1}^{n} Z_i'(y_i - X_i\widehat{\beta}) = \frac{1}{n}Z'\widehat{u} \quad (2.3)$$

The intuition behind GMM is to choose an estimator for $\beta$ that solves $\overline{g}(\widehat{\beta}) = 0$.

If the equation to be estimated is exactly identified, so that $L = K$, then we have many equations. If the $L$ moment conditions are unknown-then we have $K$ coefficients in $\widehat{\beta}$. In this case it is possible to find a $\widehat{\beta}$ that solves $\overline{g}(\beta) = 0$, and this GMM estimator is, in fact, the IV estimator.

If the equation is overidentified, however, so that $L > K$, then we have more equations than unknowns, and in general it will not be possible to find a $\widehat{\beta}$ that will set all $L$ sample moment conditions to exactly zero. In this case, we take an $L \times L$ weighing matrix $W$ and use it to construct a quadratic form in the moment conditions. This gives the GMM objective function:

$$J(\widehat{\beta}) = n\overline{g}(\widehat{\beta})' W \overline{g}(\widehat{\beta}) \qquad (2.4)$$

A GMM estimator for $\beta$ is the $\widehat{\beta}$ that minimizes $J(\widehat{\beta})$. Deriving and solving the $K$ first order conditions

$$\frac{\partial J(\widehat{\beta})}{\partial \widehat{\beta}} = 0 \qquad (2.5)$$

yields the GMM estimator:

$$\widehat{\beta}_{GMM} = (X'ZWZ'X)^{-1}X'ZWZ'y \qquad (2.6)$$

Note that the results of the minimization will be the same for weighing matrices that differ by a constant of proportionality.

### 3.2. Applying the robust instruments to the six-factor asset pricing model

#### 3.2.1. A six-factor asset pricing model

The cost of equity $E(R_i)$ for the firm $i$, is given by equation (3.1) and follows the well-known convention as in Copeland, Weston, and Shastri (2005),

$$E(R_i) - R_f = \sum_{k=1}^{n} E(\tilde{\delta}_k)\beta_{ik} \qquad (3.1)$$

where $E(\cdot)$ is the expectation operator, $\tilde{\delta}_k$ is usually an unobservable variable, and $\beta_{ik}$ is the sensitivity of stock $i$ to the unobservable variable $\tilde{\delta}_k$.

Empirically the cost of equity for the stock $i$ can be written as

$$R_i - R_f = \alpha_i + \sum_{k=1}^{n} \delta_k \beta_{ik} + \varepsilon_i \qquad (3.2)$$

where $n = 6$ for six-factor asset pricing model proposed in the study. The parameter $\alpha_i$ is the abnormal return for stock $i$ known as the Jensen (1968) performance measure, $\delta_k$ is a proxy for the unobservable variable $\tilde{\delta}_k$, and $\varepsilon_i$ is the error term. The proxy variable $\delta_k$ is defined by the matrix equation (3.3).

$$\delta = \tilde{\delta} + u \qquad (3.3)$$

$\delta$ is a matrix of the dimension $T \times n$ of the $n$ observable proxy factors that contain measurement errors and $\tilde{\delta}$ is a matrix of the dimension $T \times n$ of the factors measured with error. $u$ is a matrix of measurement errors, which we assume to be normally distributed. Substituting (3.3) into (2.2) yields (3.4).

$$R_i - R_f = \alpha_i i_T + \delta \beta_i + \varepsilon_i - u\beta_i = \alpha_i i_T + \delta \beta_i + e_i \qquad (3.4)$$

where $i_T$ is an identity vector of dimension $T \times 1$. Estimating (3.4) by OLS yields inconsistent results. This is the conventional errors-in-variable problem (Fomby, Johnson, & Hill, 1984).

#### 3.2.2. Robust instrumental variables for GMM estimation

Following Racicot and Rentz (2015) $GMM_d$ is based on the robust instrumental variable as a distance estimator. This is an extension of GMM originally proposed by Hansen (1982), the $GMM_d$ framework of the robust instrumental variable estimator may be written as,

$$\underset{\widehat{\beta}}{\operatorname{argmin}}\left\{n^{-1}\left[d'\left(Y - X\widehat{\beta}\right)\right]' W n^{-1}\left[d'\left(Y - X\widehat{\beta}\right)\right]\right\} \qquad (4.1)$$

The variables in (4.1) are defined below in (4.2) to (4.5). We start with $W$, which is a weighting matrix that can be estimated through HAC (Heteroscedasticity-consistent) estimator and $Y$ is defined as

$$y = X\beta + \varepsilon \qquad (4.2)$$

where $X$ is assumed to be an unobserved matrix of explanatory variables. The observed matrix of explanatory variables is assumed to be measured with normally distributed error, viz., $X^* = X + v$.

$\widehat{\beta}$ is defined as

$$\widehat{\beta} = \widehat{\beta}_{TSLS} = (X'P_zX)^{-1}X'PY \qquad (4.3)$$

$P_z$ is defined as standard 'predicted value maker of projection matrix' used to compute

$$P_zX = Z(Z'Z)^{-1}Z'X = Z\widehat{\theta} = \widehat{X} \qquad (4.4)$$

where Z is retrieved by optimally combining the Durbin (1954) and Pal (1980) estimators using GLS. The result is based on the Bayesian approach of Theil and Goldberger (1961). This approach for obtaining Z is implemented in equation (4.7) below in deviation form.

From (4.4) extract the matrix of residuals

$$d = X - \widehat{X} = X - P_ZX = (I - P_Z)X \qquad (4.5)$$

In (4.5) the matrix $d$ is a matrix of instruments that can be defined individually in deviation form as

$$d_{it} = x_{it} - \widehat{x}_{it} \qquad (4.6)$$

Intuitively, the variable $d_{it}$ is a filtered version of the endogenous variables. It potentially removes non-linearity that might be hidden in $x_{it}$. The $\widehat{x}_{it}$ in (4.6) are retrieved applying OLS on the z instruments.

$$\widehat{x}_{it} = \widehat{\gamma}_0 + z\widehat{\phi} \qquad (4.7)$$

The z instruments are defined as $Z = \mathbf{Z} = \{z_0, z_1, z_2\}$, where $z_0 = i_T$, $z_1 = x \bullet x$, and $z_2 = x \bullet x \bullet x - 3x[D(x'x/T)]$. The symbol $\bullet$ is the Hadamard product, $D(x'x/T) = p\lim_{T \to \infty}(x'x/T) \bullet I_n$ is a diagonal matrix, and $I_n$ is an identity matrix of dimension $n \times n$, where n is the number of independent variables, $z_1$ contains the instruments used in the Durbin (1954) estimator, and $z_2$ contains the cumulant instruments employed by Pal (1980).

It should be emphasized that the third and fourth cross sample moments are used as instruments to estimate the model parameters. It is believed that the assumption of normality is a sufficient condition for estimators to be consistent once measurement errors are purged using these third and fourth cross sample moments.

### 3.2.3. Hausman's specification test and an application to instrumental variable estimation

Following Hausman (1978), the covariance between an efficient estimator, $\mathbf{b}_E$ of a parameter vector, $\beta$, and its difference from an inefficient estimator, $b_I$, of the same parameter vector, $\mathbf{b}_E - \mathbf{b}_I$, is zero.

For our study, $\mathbf{b}_E$ is $\mathbf{b}_{OLS}$ and $\mathbf{b}_I$ is $\mathbf{b}_{IVGMM}$. By Hausman's result, can have

$$Cov[b_E, b_E - b_I] = Var[b_E] - Cov[b_E, b_I] = 0$$

or

$$Cov[b_E, b_I] = Var[b_E],$$

so,

$$Asy.Var[b_{IVGMM} - b_{OLS}] = ASy.Var[b_{IVGMM}] - Asy.Var[b_{OLS}]$$

Inserting this useful result into the Wald statistics and reverting to our empirical estimates of these quantities, we have

$$H = (b_{IVGMM} - b_{OLS})\{Est.Asy.Var[b_{IVGMM}] - Est.Asy.Var[b_{OLS}]\}^{-1}(b_{IVGMM} - b_{OLS})$$

Under the null hypothesis, we are using two different, but consistent, estimators of $\sigma^2$. If we use $s^2$ as the common estimator, then the statistics will be

$$H = \frac{d'\left[\left(\widehat{X}'\widehat{X}\right)^{-1} - (X'X)^{-1}\right]^{-1} d}{s^2} \quad (5.1)$$

## 4. Empirical results

### 4.1. Summary statistics

Panel A of Table S2 (See the Supplementary Material, available online) reports the summary statistics of factor portfolios, labor income growth rate (LBR), market factor (RM-RF), size (SMB), value (HML), profitability (RMW), and investment (CMA). The average LBR for the US is 17.61 with the reasonable deviation of 1.87. The market factor incurs low average return of 0.66 with the higher deviation of 4.42. Moreover, RMW and CMA patterns show relatively higher average returns than size and value patterns. The well-crafted fact in the asset pricing literature that the average returns on smaller stocks (twelve portfolios sorted on size-B/M, size-investment, size-profitability, and size-momentum) outweigh the bigger stocks (twelve portfolios sorted on size-B/M, size-investment, size-profitability, and size-momentum) (See panel B of Table S2, available online). Panel C of Table S2 (See the Supplementary Material, available online) shows the summary statistics of the one-hundred portfolios sorted on the intersections of size-B/M, size-investment, size-profitability, and size-momentum. The common pattern that, a decrease in average return from the small-value portfolio to the high-value portfolio is witnessed in the FF portfolios sorted on size-B/M, size-profitability, and size-momentum with an exception of the portfolios sorted on size-investment.

Panel D of Table S2 (See the Supplementary Material, available online) reports the summary statistics of Industry portfolios. The average industry return and deviation range from lowest (0.69 and 5.71) for the smoke industry to the highest (1.55 and 6.76) for books industry. Panel E of Table S2 (See the Supplementary Material, available online) reports summary statistics of index portfolios, and show all the indexes comparatively fetches the same quantum of average return and deviation ranges from lowest (5.19 and 12.36) for NASDAQ Transportation to highest (5.95 and 14.38) for Dow Jones Transportation.

### 4.2. Testing for robust instrumental variables

#### 4.2.1. Relevance test

The instance of weak instruments occurs when $((1)/(n))Z'X$ is close to zero. We move forward analogously to Olea and Pflueger (2013) who extend the work of Stock and Yogo (2005) and Stock and Watson (2015) to the general case of heteroscedasticity and autocorrelation. These authors used the conventional $F$-statistics for testing that, all the coefficients in the regression are zero.

$$x_i = z_i'\pi + v_i \quad (6)$$

Explicitly, we run regression (6) for each explanatory variable on all the instruments. According to Olea and Pflueger (2013), if the resulting $F$-statistics is smaller than 24 for all of the regressions is an indication of a potential weak instruments problem. If a least one of the $F$ values is above 24, then the instruments are robust.

Note from Table 1 that the $F$-statistics for the regression $x_2$, $x_4$, and $x_6$ are over 24. The coefficients of the instrumental variables denote the partial correlation of the instruments with the explanatory variables. The coefficient for, $z_2$ on $x_2$ to $z_6$ on

Table 1
Relevance test for robust instruments.

|  | $\alpha$ | $z_1$ | $z_2$ | $z_3$ | $z_2$ | $z_5$ | $z_6$ | F-stat |
|---|---|---|---|---|---|---|---|---|
| $x_1$ | 1.11e + 08*** | – | −239198.3 | 1631789.0 | −392156.2 | 492093.3 | −1421953 | 0.46 |
|  | 21.23 |  | −0.18 | 1.09 | −0.17 | 0.25 | −0.51 |  |
| $x_2$ | 2.04 | −0.05 | – | 0.07 | 0.29** | −0.56*** | −0.96*** | **23.86*** |
|  | 0.82 | −0.39 |  | 0.64 | 2.15 | −5.02 | −5.83 |  |
| $x_3$ | −1.88 | 0.12 | 0.03 | – | −0.04 | −0.50 | 0.09 | 7.52*** |
|  | −1.49 | 1.71 | 0.67 |  | −0.46 | −4.73 | 0.72 |  |
| $x_4$ | −0.87 | 0.04 | 0.08 | −0.02 | – | 0.29*** | 0.93*** | **48.80*** |
|  | −0.85 | 0.64 | 1.92 | −0.47 |  | 4.17 | 13.20 |  |
| $x_5$ | 0.33 | 0.01 | −0.17*** | −0.31*** | 0.30*** | – | 3.48 | 10.85*** |
|  | 0.30 | 0.15 | −4.41 | −4.42 | 3.48 |  | −1.53 |  |
| $x_6$ | 0.83 | −0.03 | −0.13*** | 0.03 | 0.44*** | −0.09 | – | **44.79*** |
|  | 1.21 | −0.73 | −5.43 | 0.73 | 13.03 | 0.127 |  |  |

The aggregate of 348 observations are used to compute the descriptive statistics. The standard error is computed using the Robust method. The values in *italic* represent *t*-statistics. According to Olea and Pflueger (2013) if *F*-statistics is less than 24.00, then it is the sign of a potential weak instrument problem, and if a least one of the *F*-statistics is more than 24.00, then the instruments are robust. The values with **bold** represent *F*-statistics that exceeds the critical value of 24.00, hence the instruments are robust. ** and *** denotes statistical significance at 5% and 1% level respectively.

$x_2$, $z_2$ on $x_4$ to $z_6$ on $x_4$, and $z_2$ on $x_6$ to $z_6$ on $x_6$, are close to 1 or above, with significant *t*-statistics. This implies instruments is correlated to its respective explanatory variables.

### 4.2.2. Exogeneity test

The instance where the instruments are uncorrelated with the error terms, viz. corr($z_{1i},\varepsilon_i$) = 0,…,corr($z_{mi},\varepsilon_i$) = 0, the instruments are exogenous. Rather than computing individual partial correlation coefficients, we regressed the instruments on the error terms as in equation (7),

$$\widehat{\varepsilon}_i = c + \gamma_1 z_{1i} + \gamma_2 z_{2i} + \gamma_3 z_{3i} + \gamma_4 z_{4i} + \gamma_5 z_{5i} + \gamma_6 z_{6i} + \xi_i \quad (7)$$

where $\widehat{\varepsilon}_i$ is the estimated residual from our proposed six-factor asset pricing regression equation (1). As in Racicot and Rentz (2015), the coefficients of this regression disaggregate the effect of each regressor with the error term. Therefore, the resulting coefficients are analogous to the partial correlation coefficients.

Table 2 reports the result of instruments exogeneity test following the regression equation (7). Note that all of the coefficients of instrumental variables in (7) are very close to zero and insignificant as their probability values (*P*-value) are substantially greater than the conventional significance levels. Conclusively, it is evident that instruments are indeed exogenous.

## 5. Estimating the six-factor asset pricing model with specification errors

The OLS and IVGMM estimates of the parameters of the new six-factor asset pricing model appear in panel A for the size-B/M portfolio, panel B for size-investment portfolio, panel C for size-momentum portfolio, and panel D for index portfolios of Table 3. In panel A of Table 3 for six portfolios sorted on size-B/M, the coefficient of LBR is significant for five portfolios using OLS whereas three using IVGMM. Further, the coefficient of RM-RF is significant for all the six portfolios using both OLS and IVGMM. Similarly, the coefficient of SMB factor is significant for five of the six portfolios using both OLS and IVGMM. The coefficients of HML and CMA factors are significant for all the six portfolios whereas the coefficient of RMW is significant for five portfolios using OLS. The adjusted $R^2$ (coefficient of variation) of the six-factor model is higher using OLS (0.97) than IVGMM (0.86) approach. The HML, RMW, and CMA are a set of instruments used while estimating the parameters of the new six-factor model using IVGMM approach. We argue in line with the findings of Roy and Shijin (2018) that human capital component subsumes the predictive power of value strategy and likewise, size factor subsumes the largest portion of predictability of investment and profitability strategies respectively. Henceforth, we primarily discuss and concentrate the results, that comprised of the human capital component (LBR), market factor (RM-RF), and the size factor (HML) respectively. Further, panel A of Table S3 (See the Supplementary Material, available online) shows the results of overidentification, weak instrumental, and Hausman tests, for the six portfolios sorted on size-B/M. The overidentification and weak instrumental tests results show the instruments used in IVGMM are valid and Hausman specification test specifies the IVGMM estimates are consistent for five of the six portfolios. This suggests that LBR contain the measurement error

Table 2
Exogeneity test for robust instruments.

|  | $\alpha$ | $z_1$ | $z_2$ | $z_3$ | $z_2$ | $z_5$ | $z_6$ |
|---|---|---|---|---|---|---|---|
| Coef | 2.11E-14 | −1.23E-15 | 6.28E-16 | 2.32E-16 | −7.12E-16 | 8.76E-16 | 1.39E-15 |
| P-value | 1.00 | 1.00 | 1.00 | 1.00 | 1.00 | 1.00 | 1.00 |
| $R^2$ | – | – | – | – | – | – | – |

Note: An aggregate of 377 observations are used to estimate the descriptive statistics. The data reported in this table are from FF portfolio sorted on the intersection of size-B/M (for an instance).

Table 3
OLS vs IVGMM estimation method for the six-factor model.

| Panel A: Bivariate sorted Size-B/M portfolio | | | | | | | | | | |
|---|---|---|---|---|---|---|---|---|---|---|
| No. of portfolios | | a | LBR | RM-RF | SMB | HML | RMW | CMA | $\overline{R}^2$ | DW |
| Six | OLS | 1.67 | −0.21 | 1.02 | 0.41 | 0.23 | 0.03 | 0.02 | 0.970 | 1.64 |
| | t-mean | 3.46 | −3.75 | 95.38 | 30.06 | 9.13 | 1.68 | −0.45 | | |
| | t-min | 0.15 | −5.41 | 62.57 | −10.73 | −18.29 | −11.53 | −11.53 | | |
| | t-max | 6.56 | −1.28 | 123.36 | 74.05 | 35.26 | 10.19 | 4.21 | | |
| | No. of signif. portfolios | 5 | 5 | 6 | 5 | 6 | 5 | 6 | | |
| | IVGMM | 1.36 | −0.17 | 0.98 | 0.28 | | | | 0.869 | 1.50 |
| | z-mean | 2.59 | −1.70 | 25.13 | 3.77 | | | | | |
| | z-min | 0.58 | −2.82 | 18.82 | −4.48 | | | | | |
| | z-max | 5.05 | 0.02 | 32.68 | 16.05 | | | | | |
| | No. of signif. portfolios | 3 | 3 | 6 | 5 | | | | | |
| **Panel B: Bivariate sorted Size-investment portfolio** | | | | | | | | | | |
| Six | OLS | 1.66 | −0.08 | 1.01 | 0.40 | 0.05 | −0.01 | 0.09 | 0.978 | 1.65 |
| | t-mean | 4.51 | −3.70 | 101.25 | 29.53 | 2.92 | −0.49 | 3.37 | | |
| | t-min | 3.51 | −4.43 | 89.98 | −8.71 | −0.99 | −16.16 | −19.18 | | |
| | t-max | 5.48 | −2.78 | 114.19 | 72.78 | 12.19 | 8.37 | 21.68 | | |
| | No. of signif. portfolios | 6 | 6 | 6 | 6 | 3 | 6 | 5 | | |
| | IVGMM | 1.72 | −0.07 | 0.98 | 0.42 | | | | 0.921 | 1.66 |
| | z-mean | 2.85 | −2.07 | 33.61 | 3.80 | | | | | |
| | z-min | 1.19 | −3.50 | 0.02 | −6.68 | | | | | |
| | z-max | 4.35 | −0.88 | 53.71 | 14.14 | | | | | |
| | No. of signif. portfolios | 5 | 3 | 6 | 4 | | | | | |
| **Panel C: Bivariate sorted Size-momentum portfolio** | | | | | | | | | | |
| Six | OLS | 1.48 | −0.23 | 1.03 | 0.39 | 0.22 | 0.02 | −0.13 | 0.868 | 1.63 |
| | t-mean | 1.53 | −1.43 | 44.57 | 12.21 | 4.09 | 2.36 | −0.70 | | |
| | t-min | −0.52 | −3.46 | 24.57 | −5.68 | −3.63 | −4.16 | −4.28 | | |
| | t-max | 4.26 | 0.89 | 65.05 | 35.12 | 12.32 | 7.97 | 1.90 | | |
| | No. of signif. portfolios | 2 | 2 | 6 | 4 | 6 | 3 | 3 | | |
| | IVGMM | 1.38 | −0.06 | 1.02 | 0.28 | | | | 0.805 | 1.68 |
| | z-mean | 1.33 | −0.93 | 21.33 | 0.01 | | | | | |
| | z-min | −0.04 | −2.20 | 11.07 | −7.90 | | | | | |
| | z-max | 3.10 | 0.48 | 37.26 | 5.18 | | | | | |
| | No. of signif. portfolios | 3 | 2 | 5 | 3 | | | | | |
| **Panel D: Index portfolios** | | | | | | | | | | |
| Five | OLS | 38.38 | −1.88 | 1.02 | 0.22 | 0.38 | −0.14 | 0.33 | 0.149 | 0.18 |
| | t-mean | 5.33 | −3.63 | 5.27 | 0.81 | 1.09 | 0.16 | 0.86 | | |
| | t-min | 4.51 | −5.10 | 4.05 | −1.71 | 0.23 | −3.87 | −0.61 | | |
| | t-max | 5.72 | 0.38 | 6.31 | 2.70 | 2.36 | 1.51 | 1.76 | | |
| | No. of signif. portfolios | 5 | 5 | 5 | 2 | 1 | 1 | 0 | | |
| | IVGMM | 38.91 | −1.85 | 0.94 | 0.45 | | | | 0.099 | 0.75 |
| | z-mean | 2.95 | −2.40 | 6.17 | −0.15 | | | | | |
| | z-min | 1.90 | −2.89 | 3.10 | −3.71 | | | | | |
| | z-max | 3.52 | −1.60 | 7.90 | 4.41 | | | | | |
| | No. of signif. portfolios | 4 | 4 | 4 | 2 | | | | | |

The results appearing in this table are the averages of the six-portfolio each, sorted on the intersections of, size-B/M in panel A, size-investment in panel B, size-momentum in panel C, and the averages of index portfolios in panel D, respectively. t-statistics are in italics and are HAC (Newey & West, 1987) corrected for IVGMM. The number of significant portfolios at 5% level are labelled by No. of signif. Portfolios. The Durbin-Watson statistics is indicated by $DW$ and $\overline{R}^2$ is the adjusted $R$ squared. While using IVGMM for estimating the parameters of the six-factor model with specification errors, SMB acts as endogenous variable and HML, RMW, CMA is used as instruments, whereas RM-RF and LBR are considered as the exogenous variable.

since the coefficient of LBR is significant for five of the six portfolios using OLS and three using IVGMM.

In panel B of Table 3 for six portfolios sorted on size-investment, the coefficient of LBR is significant, for all the six portfolios using OLS and three using IVGMM. The coefficient of RM-RF is significant for all the six portfolios using both OLS and IVGMM. The coefficient of SMB factor is significant for all the six portfolios using OLS and four using IVGMM. The adjusted $R^2$ for the six-factor model is higher using OLS (0.97) than IVGMM (0.92). Further, panel B of Table S3 (See the Supplementary Material, available online) consists of the robustness test results that show the instruments used in IVGMM are valid and the estimates are consistent for all the portfolios sorted on size-investment. Since the coefficient of LBR is significant for all the portfolios using OLS and three using IVGMM. This suggests that LBR contain the measurement error while estimating the parameters of the six-factor model using OLS indicating that IVGMM approach should be used in the first place.

In panel C of Table 3 for six portfolios sorted on size-momentum, the coefficient of LBR is significant for two portfolios using both OLS and IVGMM. The coefficient of RM-RF is significant for all the six portfolios using OLS whereas five portfolios are significant using IVGMM. The coefficient of SMB factor is significant for four portfolios using OLS and three using IVGMM. The adjusted $R^2$ of the six-factor model is higher using OLS (0.86) than IVGMM (0.80). Panel C of Table S3 (See the Supplementary Material, available online) reports the robustness test results, which indicates that the instruments used in IVGMM are robust and the estimates are consistent for all the portfolios. Moreover, the coefficients of RM-RF and SMB factor are significant for, six and four portfolios respectively using OLS and five and three using IVGMM. This suggests that RM-RM and SMB factor contain the measurement error while estimating the parameters of the six-factor model using OLS and concurrently indicates that the IVGMM approach is the robust estimator. The result is consistent with Racicot and Rentz (2015) that SMB factor may contain measurement errors.

In panel D of Table 3 for five indexes (TRAN, CTRN, CINS, CUTL, CBNK) portfolios, the coefficient of LBR and SMB are significant for five portfolios using OLS approach and four using IVGMM approach. The coefficient of SMB is significant for two portfolios using both OLS and IVGMM approach. The adjusted $R^2$ of the six-factor model is relatively higher using OLS (0.14) approach than IVGMM (0.09) approach. Concurrently, panel D of Table S3 (See the Supplementary Material, available online) reports the robustness test results, which suggest the instruments used in IVGMM estimation are valid and the estimates are consistent for the index portfolios. Parallely, the coefficient of LBR and RM-RF are significant for five of the six index portfolios using OLS and four portfolios using IVGMM approach. This suggests that LBR and RM-RF contain the measurement error while estimating the parameters of the six-factor model using OLS and indicates that the IVGMM approach is the robust estimator of the six-factor model. The result is in line with Racicot and Rentz (2015, 2016b) that the SMB factor contains measurement errors.

Summarily, the portfolios sorted on size-B/M, size-investment, size-momentum, and index portfolios, are employed to estimate the parameters of the new six-factor asset pricing model using OLS and IVGMM approaches. The robustness test results indicate that the parameter estimates of six-factor model following OLS approach contain measurement error while IVGMM approach emerged as the better estimator. The IVGMM is robust in estimating the parameters of multi-factor asset pricing model. The OLS may be overstating the significance of the regressions. The human capital component of six-factor model consistently priced the variation in return on variant asset class used in the present study. Further, we find favorable evidence that ensures market factor consistently price the variations in return on all the asset class. Simultaneously, it is evident that the size factor enables to explain the risk and return relationship quite well in our new six-factor model for the variant asset class.

In the successive section, we further check the robustness of the parameter estimation of new six-factor asset pricing model using OLS and IVGMM approach for the variant asset class. We used four sets of twenty-five portfolios each, sorted on the intersections of size-B/M, size-investment, size-profitability, size-momentum, and thirty industry portfolios.

## 6. Empirical performance of six-factor model: OLS vs IVGMM

Table 4 reports the result of OLS and IVGMM estimates of the six-factor model for the twenty-five portfolio each, sorted on, the size-B/M in panel A, size-investment in panel B, size-profitability in panel C, and size-momentum in panel D, respectively, and thirty industry portfolios in panel E. For twenty-five portfolios sorted on size-B/M in panel A, the LBR coefficient is significant for thirteen portfolios using OLS while twenty-three portfolios using IVGMM approach. The coefficient of RM-RF is significant for nineteen portfolios using OLS whereas all the portfolios are significant using IVGMM approach. However, SMB coefficient is significant for all the portfolios using OLS and twenty-two using IVGMM. The adjusted $R^2$ for the six-factor model using OLS is 0.91. Table S4 (See the Supplementary Material, available online) shows the comprehensive results of OLS and IVGMM estimates that indicates the instruments used in IVGMM are valid and the estimates are robust for all the portfolios. Panel A of Table 4 and Table S4 (See the Supplementary Material, available online) shows the coefficient of SMB factor is significant for all the portfolios sorted on size-B/M using OLS and twenty-two of the twenty-five portfolios using IVGMM. This suggests that the SMB factor contains significant measurement errors. The results are consistent with Racicot and Rentz (2015).

Panel B of Table 4 reports the result for twenty-five portfolios sorted on size-investment. The LBR coefficient is significant for ten portfolios using OLS whereas twenty portfolios using IVGMM approach. The coefficient of RM-RF is significant for twenty-three portfolios using OLS while significant for all the portfolios using IVGMM approach. Concurrently, the SMB factor coefficient is significant for twenty-four portfolios using OLS while it is significant for all the portfolios using IVGMM. The adjusted $R^2$ for the six-factor model using OLS is 0.92. Table S5 (See the Supplementary Material, available online) reports the comprehensive results of OLS and IVGMM estimates, which indicates the instruments used in IVGMM are robust and the IVGMM estimates are consistent for all the portfolios.

Panel C of Table 4 reports the result of twenty-five portfolios sorted on size-profitability, where the LBR coefficient is significant for fourteen portfolios using OLS and twenty using

Table 4
OLS vs IVGMM estimation method for the six-factor model (FF (5 × 5 sort) and industry portfolios).

Panel A: 5 × 5 sorted Size-B/M portfolio

| No. of portfolios | | a | LBR | RM-RF | SMB | HML | RMW | CMA | $\overline{R}^2$ | DW |
|---|---|---|---|---|---|---|---|---|---|---|
| Twenty-five | OLS | 1.64 | 0.08 | 1.02 | 0.52 | 0.23 | 0.04 | 0.03 | 0.912 | 1.70 |
| | *t*-mean | *2.13* | *−1.79* | *47.57* | *17.25* | *5.00* | *1.33* | *0.76* | | |
| | *t*-min | *0.52* | *−3.68* | *32.17* | *−11.52* | *−14.34* | *−11.73* | *−5.48* | | |
| | *t*-max | *4.43* | *−0.25* | *74.14* | *36.10* | *20.96* | *7.96* | *3.77* | | |
| | No. of signif. portfolios | 14 | 13 | 19 | 25 | 23 | 16 | 9 | | |
| | IVGMM | | 0.02 | 0.98 | 0.48 | | | | | |
| | *z*-mean | | *3.25* | *30.86* | *13.20* | | | | | |
| | *z*-min | | *−4.22* | *0.03* | *−7.83* | | | | | |
| | *z*-max | | *6.50* | *61.79* | *56.23* | | | | | |
| | No. of signif. portfolios | | 23 | 25 | 22 | | | | | |

Panel B: 5 × 5 sorted Size-investment portfolio

| Twenty-five | OLS | 1.55 | −0.07 | 1.01 | 0.53 | 0.09 | 0.01 | 0.09 | 0.924 | 1.75 |
|---|---|---|---|---|---|---|---|---|---|---|
| | No. of signif. portfolios | 12 | 10 | 23 | 24 | 14 | 18 | 19 | | |
| | IVGMM | | 0.02 | 1.01 | 0.24 | | | | | |
| | No. of signif. portfolios | | 20 | 25 | 25 | | | | | |

Panel C: 5 × 5 sorted Size-profitability portfolio

| Twenty-five | OLS | 1.56 | −0.07 | 1.02 | 0.52 | 0.11 | −0.03 | 0.01 | 0.96 | 1.64 |
|---|---|---|---|---|---|---|---|---|---|---|
| | No. of signif. portfolios | 14 | 14 | 25 | 25 | 19 | 21 | 14 | | |
| | IVGMM | | 0.08 | 0.94 | 0.49 | | | | | |
| | No. of signif. portfolios | | 20 | 25 | 25 | | | | | |

Panel D: 5 × 5 sorted Size-momentum portfolio

| Twenty-five | OLS | 1.15 | −0.06 | 1.04 | 0.49 | 0.25 | 0.05 | −0.11 | 0.838 | 1.67 |
|---|---|---|---|---|---|---|---|---|---|---|
| | No. of signif. portfolios | 6 | 6 | 24 | 23 | 23 | 19 | 9 | | |
| | IVGMM | | 0.03 | 0.97 | 0.47 | | | | | |
| | No. of signif. portfolios | | 24 | 25 | 23 | | | | | |

Panel E: Industry portfolios

| Thirty | OLS | 1.60 | −0.08 | 1.07 | 0.17 | 0.13 | 0.29 | 0.13 | 0.626 | 1.83 |
|---|---|---|---|---|---|---|---|---|---|---|
| | No. of signif. portfolios | 3 | 3 | 30 | 20 | 22 | 21 | 11 | | |
| | IVGMM | | 0.02 | 1.01 | 0.073 | | | | | |
| | No. of signif. portfolios | | 17 | 30 | 19 | | | | | |

The results appearing in this table are the averages of the twenty-five portfolio each, sorted on the intersections of, size-B/M in panel A, size-investment in panel B, size-momentum in panel C, and the averages of thirty industry portfolios in panel D, respectively. *t*-statistics are in italics and are HAC (Newey & West, 1987) corrected for IVGMM. The number of significant portfolios at 5% level are labelled as No. of signif. portfolios. The Durbin-Watson statistics is indicated by *DW* and $\overline{R}^2$ is the adjusted *R* squared. While using IVGMM for estimating the parameters of the six-factor model with specification errors, SMB acts as endogenous variable and HML, RMW, CMA is used as instruments, whereas RM-RF and LBR are considered as the exogenous variable. The comprehensive results of, Panel A, B, C, D, and E, are reported in Table S4, S5, S6, S7, and S8, respectively.

IVGMM. The coefficients of RM-RF and SMB factor are significant for all the portfolios using both OLS and IVGMM. The adjusted $R^2$ for the six-factor model using OLS is 0.96. Table S6 (See the Supplementary Material, available online) shows the comprehensive results of OLS and IVGMM estimates, which indicates that the instruments used in IVGMM are valid and the estimates are robust for all the portfolios.

Panel D of Table 4 reports the result of twenty-five portfolios sorted on size-momentum, where the LBR coefficient is significant for three portfolios using OLS and twenty-four using IVGMM. The coefficient of RM-RF is significant for twenty-four portfolios using OLS albeit all the portfolios are significant using IVGMM approach. Following, SMB coefficient is significant for twenty-three portfolios using both OLS and IVGMM. The adjusted $R^2$ for the six-factor model using OLS is 0.83. The tests result in Table S7 (See the Supplementary Material, available online) show that the instruments used in IVGMM are valid and the estimates are robust for all the portfolios.

Panel E of Table 4 reports the result of thirty industry portfolios, where the LBR coefficient is significant for three industry portfolios using OLS and seventeen using IVGMM approach. The coefficient of RM-RF is significant for all the thirty index portfolios using both OLS and IVGMM. Following, SMB coefficient is significant for twenty industry portfolios using OLS and nineteen using IVGMM. The adjusted $R^2$ for the six-factor model using OLS is 0.62. The robustness test results in Table S8 (See the Supplementary Material, available online) show that the IVGMM estimates are consistent and robust for all the index portfolios.

Briefly, the four sets of twenty-five portfolios each sorted on, size-B/M, size-investment, size-profitability, and size-momentum, and thirty industry portfolios are used to explain the variations in return predictability. OLS and IVGMM approaches are used to estimate the parameters of the six-factor asset pricing model. The robustness test results indicate that the parameters estimate of the six-factor model using OLS consists of measurement errors. Further, the IVGMM estimate enhances the performance of the new six-factor model and hence the IVGMM is the robust instrument approach in estimating the parameters of multi-factor asset pricing models. Moreover, the OLS may be overstating the significance of the regressions. Further, the human capital component of six-factor model consistently priced the variation in return on variant asset class used in the study.

Harvey et al. (2016) argued that unless the *t*-ratio for a factor is more than 3.00, any claimed empirical finding for a factor is likely to be false and is the result of data mining. Further, Cochrane (2011) casts the doubt about the importance of the plethora of such factors discovered. Motivated by this argument, we check the performance and persistence of human capital component of the six-factor asset pricing model using the criteria of *t*-ratio. This test will further hold our argument that the dynamic human capital component of the six-factor asset pricing model identified as a standalone predictor of the asset returns. The results are discussed in the successive section.

## 7. Dynamic human capital

We assess the *t*-ratio of the human capital component (LBR) of each IVGMM estimates of the six-factor asset pricing model for the variant asset class used in the study. Our claim that the dynamic human capital component is a standalone predictor in the six-factor asset pricing model will further justify and strengthen if the *t*-ratio of the human capital component in the respective regression using IVGMM estimation exceeds the criterion limit of 3.00. Table 5 shows the *t*-ratio approximation of the human capital coefficients using IVGMM approach for the variant portfolios. Table 5 reports the *t*-ratio of the human capital coefficients that clear the cutoff of *t*-ratio greater than 3.00. The robustness test results show that the instruments used in IVGMM are valid and the estimates are robust for all the portfolios.

Table 5 shows the human capital component successively price the variations in return on the aggregate of eighty-three variant portfolios. The variant portfolios include three sets of six portfolios each, sorted on the intersections of size-B/M, size-investment, and size-momentum. Five index portfolios and four sets of twenty-five portfolios each sorted on the intersections of size-B/M, size-investment, size-profitability, and size-momentum, and, thirty industry portfolios. The sensitivity of the human capital component (LBR coefficient) reported in Table 5 exceeding the *t*-ratio of 3.00, is statistically and economically significant at all the conventional levels. Hence,

Table 5
*t*-ratio approximation of the human capital (LBR) component of a six-factor model.

| Sl. No. | Portfolio | Portfolio type | LBR Coefficient | *t*-ratio (±3) |
|---|---|---|---|---|
| 1 | Size-Investment (2 × 3) | SLoINV | −0.14*** | −3.16 |
| 2 | Size-Investment (2 × 3) | BLoINV | −0.13*** | −3.5 |
| 3 | Size-B/M (5 × 5) | Small/LoBM | −0.03*** | −4.22 |
| 4 | Size-B/M (5 × 5) | Small/4 | 0.03*** | 6.5 |
| 5 | Size-B/M (5 × 5) | Small/HiBM | 0.03*** | 4.21 |
| 6 | Size-B/M (5 × 5) | 2/2 | 0.02*** | 4.16 |
| 7 | Size-B/M (5 × 5) | 2/3 | 0.03*** | 6.1 |
| 8 | Size-B/M (5 × 5) | 2/4 | 0.03*** | 4.4 |
| 9 | Size-B/M (5 × 5) | 2/HiBM | 0.03*** | 4.16 |
| 10 | Size-B/M (5 × 5) | 3/2 | 0.01*** | 4.05 |
| 11 | Size-B/M (5 × 5) | 3/3 | 0.02*** | 4.29 |
| 12 | Size-B/M (5 × 5) | 3/4 | 0.03*** | 4.36 |
| 13 | Size-B/M (5 × 5) | 3/HiBM | 0.04*** | 3.92 |
| 14 | Size-B/M (5 × 5) | 4/2 | 0.02*** | 3.31 |
| 15 | Size-B/M (5 × 5) | 4/4 | 0.03*** | 5.26 |
| 16 | Size-B/M (5 × 5) | 4/HiBM | 0.03*** | 3.13 |
| 17 | Size-B/M (5 × 5) | Big/2 | 0.02*** | 4.97 |
| 18 | Size-B/M (5 × 5) | Big/3 | 0.02*** | 3.95 |
| 19 | Size-B/M (5 × 5) | Big/HiBM | 0.03*** | 3.46 |
| 20 | Size-Investment (5 × 5) | Small/LoINV | 0.02*** | 3.25 |
| 21 | Size-Investment (5 × 5) | Small/2 | 0.03*** | 5.76 |
| 22 | Size-Investment (5 × 5) | Small/3 | 0.03*** | 7.86 |
| 23 | Size-Investment (5 × 5) | Small/4 | 0.02*** | 4.24 |
| 24 | Size-Investment (5 × 5) | Small/HiINV | 0.02*** | −3.42 |
| 25 | Size-Investment (5 × 5) | 2/2 | 0.04*** | 4.88 |
| 26 | Size-Investment (5 × 5) | 2/3 | 0.04*** | 6.47 |
| 27 | Size-Investment (5 × 5) | 2/4 | 0.03*** | 6.06 |
| 28 | Size-Investment (5 × 5) | 3/LoINV | 0.03*** | 3.99 |
| 29 | Size-Investment (5 × 5) | 3/2 | 0.03*** | 6.3 |
| 30 | Size-Investment (5 × 5) | 3/3 | 0.03*** | 5.03 |
| 31 | Size-Investment (5 × 5) | 3/4 | 0.02*** | 3.96 |
| 32 | Size-Investment (5 × 5) | 4/LoINV | 0.02*** | 4.12 |
| 33 | Size-Investment (5 × 5) | 4/2 | 0.03*** | 3.37 |
| 34 | Size-Investment (5 × 5) | 4/3 | 0.03*** | 5.84 |
| 35 | Size-Investment (5 × 5) | 4/4 | 0.02*** | 4.49 |
| 36 | Size-Investment (5 × 5) | Big/LoINV | 0.03*** | 4.5 |
| 37 | Size-Investment (5 × 5) | Big/3 | 0.02*** | 6.67 |
| 38 | Size-Investment (5 × 5) | Big/4 | 0.02*** | 4.48 |
| 39 | Size-OP (5 × 5) | LoBM/2 | 0.03*** | 5.52 |
| 40 | Size-OP (5 × 5) | LoBM/3 | 0.03*** | 5.31 |
| 41 | Size-OP (5 × 5) | LoBM/4 | 0.03*** | 6.54 |
| 42 | Size-OP (5 × 5) | LoBM/HiOP | 0.02*** | 3.87 |
| 43 | Size-OP (5 × 5) | 2/2 | 0.02*** | 3.15 |
| 44 | Size-OP (5 × 5) | 2/3 | 0.03*** | 7.94 |
| 45 | Size-OP (5 × 5) | 2/4 | 0.03*** | 4.65 |
| 46 | Size-OP (5 × 5) | 2/HiOP | 0.03*** | 4.89 |
| 47 | Size-OP (5 × 5) | 3/2 | 0.02*** | 4.33 |
| 48 | Size-OP (5 × 5) | 3/3 | 0.02*** | 3.93 |
| 49 | Size-OP (5 × 5) | 3/HiOP | 0.02*** | 3.95 |
| 50 | Size-OP (5 × 5) | 4/3 | 0.02*** | 4.74 |
| 51 | Size-OP (5 × 5) | 4/4 | 0.02*** | 3.53 |
| 52 | Size-OP (5 × 5) | 4/HiOP | 0.02*** | 5.55 |
| 53 | Size-OP (5 × 5) | HiBM/4 | 0.02*** | 5.84 |
| 54 | Size-OP (5 × 5) | HiBM/HiOP | 0.03*** | 4.53 |
| 55 | Size-Momentum (5 × 5) | Small/Low | −0.04*** | −4.06 |
| 56 | Size-Momentum (5 × 5) | Small/3 | 0.03*** | 5.44 |
| 57 | Size-Momentum (5 × 5) | Small/4 | 0.04*** | 4.55 |
| 58 | Size-Momentum (5 × 5) | Small/High | 0.05*** | 4.09 |
| 59 | Size-Momentum (5 × 5) | 2/Low | −0.03*** | −4.05 |
| 60 | Size-Momentum (5 × 5) | 2/2 | 0.01*** | 3.69 |
| 61 | Size-Momentum (5 × 5) | 2/3 | 0.03*** | 5.67 |

Table 5 (*continued*)

| Sl. No. | Portfolio | Portfolio type | LBR Coefficient | t-ratio (±3) |
|---|---|---|---|---|
| 62 | Size-Momentum (5 × 5) | 2/4 | 0.03*** | **5.81** |
| 63 | Size-Momentum (5 × 5) | 2/High | 0.04*** | **4.02** |
| 64 | Size-Momentum (5 × 5) | 3/3 | 0.03*** | **5.13** |
| 65 | Size-Momentum (5 × 5) | 3/4 | 0.02*** | **3.94** |
| 66 | Size-Momentum (5 × 5) | 3/High | 0.04*** | **3.39** |
| 67 | Size-Momentum (5 × 5) | 4/3 | 0.03*** | **4.76** |
| 68 | Size-Momentum (5 × 5) | 4/4 | 0.02*** | **4.3** |
| 69 | Size-Momentum (5 × 5) | 4/High | 0.04*** | **4.13** |
| 70 | Size-Momentum (5 × 5) | Big/3 | 0.01*** | **4.16** |
| 71 | Size-Momentum (5 × 5) | Big/4 | 0.02*** | **5.13** |
| 72 | Size-Momentum (5 × 5) | Big/High | 0.03*** | **4.63** |
| 73 | Industry | Food | 0.03*** | **3.04** |
| 74 | Industry | Beer | 0.04*** | **5.34** |
| 75 | Industry | Smoke | 0.64*** | **3.22** |
| 76 | Industry | Hshld | 0.03*** | **5.26** |
| 77 | Industry | Hlth | 0.04*** | **4.09** |
| 78 | Industry | Chems | 0.02*** | **3.16** |
| 79 | Industry | Util | 0.03*** | **4.94** |
| 80 | Industry | Trans | 0.02*** | **3.34** |
| 81 | Industry | Rtail | 0.03*** | **3.15** |
| 82 | Industry | Meals | 0.03*** | **3.54** |
| 83 | Industry | Fin | 0.02*** | **3.03** |

The table reports the *t*-ratio of LBR coefficients from the regressions (using the IVGMM) of the six-factor model for the variant portfolios considered in the study. We report the *t*-ratio of LBR coefficients, which is more than 3.00. We ignore the sign of the *t*-ratio since it only serves to assess the directional relationship between the variables. ** and *** indicates statistical significance at 5% and 1% level respectively. We report the *t*-ratio of the respective LBR coefficient using IVGMM estimates for the portfolios. The values of *t*-ratio of LBR is marked with **bold** face.

the predictive power of dynamic human capital component in asset return predictability enables as the strong state factor variable in the six-factor asset pricing framework. Further, these favorable empirical results, which is statistically and economically significant and valid, justify the credibility of six-factor asset pricing model.

## 8. Summary and discussion

Fama and French (2015) proposed a five-factor model that captures the size, value, profitability, and investment patterns in average stock returns. However, the five-factor model was a failure because of its inability to capture the low average returns on small stocks (Fama & French, 2015b, 2016, 2017). Successively, the failure of FF five-factor model acts as the catalyst to the debate among the financial economist on the issue of the search of the robust equilibrium asset pricing model that best captures the variations in stock returns. Campbell (1996) opines that adding human capital component along with common factors may enhance the performance of multi-factor asset pricing model in returns predictability. Rational agents maximize the expected utility of their lifetime consumption to form the mix of the optimum portfolio. The investment opportunities in both tradable and non-tradable assets force investors to invest part of their wealth in the non-tradable assets. Undoubtedly, the human capital component shared the major chunk of the wealth for almost entire agents. The study reveals that human capital component along with aggregate market is, in fact, the significant predictors of returns predictability.

Thus, we propose a new six-factor asset pricing model by adding the human capital component to the FF five-factor model to explain the variations in portfolio returns. The state variables of the equilibrium six-factor model include market factor, size, value, profitability, investment, alongside human capital. We tested the performance of six-factor model proposed in the study on the US data. We considered variant, FF portfolios, index portfolios, and industry portfolios for estimating the parameters of the six-factor model using OLS and GMM-based robust instrumental variables approaches.

The human capital, market factor, and size factor coefficients are significant for eight, seventeen and twelve portfolios respectively of first set of eighteen portfolios using IVGMM approach. In the second set of five index portfolios, the human capital and market factors are significant for four and the size factor for two portfolios respectively using IVGMM approach. The coefficient of variation of the six-factor model using OLS is relatively higher than the IVGMM. For the third set of one-hundred FF portfolios, the human capital, market, and size factors are significant for forty-three, ninety-one, and ninety-seven portfolios respectively using OLS while eighty-seven, one-hundred, and ninety-five portfolios respectively using IVGMM approach. Subsequently, for the fourth set of thirty industry portfolios, the human capital, market, and size factors are significant for three, thirty, and twenty industry portfolios respectively using OLS, while seventeen, thirty, and nineteen respectively using IVGMM approach. The human capital, market, and the size factors contain measurement errors while estimating using OLS for the first and second sets of variant portfolios. Furthermore, this suggests that OLS may be overstating the significance of the regressions.

The primary contribution of the present study to the asset pricing literature is the new six-factor asset pricing model where the human capital component is a risk factor in returns predictability. The study bridges the gap in the existing asset pricing literature of the human capital and the common risk factors by proposing a six-factor asset pricing model to better understand the risk and return relationship. Further, the study unlocks the causes that triggered the failure of FF five-factor model. Our results show that the FF five-factor model suffers from the problem of measurement errors in the variables and mostly because of estimation problem that triggered due to the model misspecification.

## 9. Conclusion

The present study proposed an equilibrium six-factor asset-pricing model by adding a human capital component with Fama & French (2015) five-factor model, which consists of the market factor, size, value, profitability, and investment to explain the variations in asset returns. We employ the four sets of variant portfolios to test the six-factor asset pricing model. The parameter estimates of the six-factor model for portfolios considered in the study, using OLS show that the human capital component equally shares the predictive power alongside the rest of the factors in the model. Concurrently, the parameter estimates of the six-factor model for the variant portfolios using IVGMM technique indicates that the human capital significantly prices the variations in return predictability alongside the rest of the factors in the model. Moreover, the relevance, exogeneity, overidentifying restrictions, and the Hausman's specification, tests indicates that the parameter estimates of the six-factor model for the four sets of variant portfolios using IVGMM is robust and outperforms the estimations obtained from OLS approach. This suggests that OLS overstates the significance of the regressions.

Furthermore, we assess the *t*-ratio of the human capital component of each IVGMM estimates of the six-factor asset pricing model for the four sets of variant portfolios used in the study. The *t*-ratio of the human capital (LBR coefficients) of the eighty-three IVGMM estimates of the six-factor model is more than 3.00 indicating the empirical success of the six-factor asset pricing model.

**Conflict of interest statement**

The authors declares that there is no conflict of interest.

**Acknowledgement**

We thank the anonymous referees and the editor for their valuable comments that improved the manuscript substantially. We thank Prachand Narayan, Assistant Professor of English, Rajiv Gandhi University, Arunachal Pradesh for language editing and reading the manuscript. The corresponding author acknowledges that the present work is the part of unpublished doctoral dissertation 'Essays in Asset pricing, Human capital, Volatility, and Income inequality: The World evidence'.

**Appendix A. Supplementary data**

Supplementary data related to this article can be found at https://doi.org/10.1016/j.bir.2018.02.001